\documentclass[hyper,preprint,a4paper,10pt]{JHEP3} % 10pt is ignored!
\usepackage{epsfig}
\usepackage{latexsym}
\usepackage{graphicx}
\usepackage{amsmath}
\usepackage{amsfonts}   % if you want the fonts
\usepackage{amssymb}    % if you want extra symbols
\newcommand\fverb{\setbox\pippobox=\hbox\bgroup\verb}
\newcommand\fverbdo{\egroup\medskip\noindent%
  \fbox{\unhbox\pippobox}\ }
\newcommand\fverbit{\egroup\item[\fbox{\unhbox\pippobox}]}
\newbox\pippobox

\newcommand{\be}{\begin{equation}}
\newcommand{\ben}{\begin{subequations}}
\newcommand{\een}{\end{subequations}}
\newcommand{\beq}{\begin{eqalignno}}
\newcommand{\eeq}{\end{eqalignno}}
\newcommand{\ee}{\end{equation}}

\newcommand{\bea}{\begin{eqnarray}}
\newcommand{\eea}{\end{eqnarray}}

\newcommand{\newc}{\newcommand}
\newc{\lcal}{\int {\cal L}dt}
\newc{\LSP}{{\chi^0_1}}
\newc{\stauR}{{\tilde \tau_R}}
\newc{\stau}{{\tilde \tau_1}}
\newc{\mstop}{m_{\tilde{t}}}
\newc{\mHpm}{m_{H^\pm}}
\newc{\ie}{{\it i.e.}}
\newc{\etal}{{\it et al.}}
\newc{\eg}{{\it e.g.}}
\newc{\kev}{\hbox{\rm\,keV}}
\newc{\mev}{\hbox{\rm\,MeV}}
\newc{\gev}{\hbox{\rm\,GeV}}
\newc{\tev}{\hbox{\rm\,TeV}}
\newc{\xpb}{\hbox{\rm\, pb}}
\newc{\xfb}{\hbox{\rm\, fb}}

\newc{\mtop}{m_t}
\newc{\mbot}{m_b}
\newc{\mz}{m_Z}
\newc{\mw}{M_W}
\newc{\alphasmz}{\alpha_s(m_Z^2)}
\newc{\swsq}{\sin^2\theta_W}
\newc{\tw}{\tan\theta_W}
\newc{\cw}{\cos\theta_W}
\newc{\sw}{\sin\theta_W}
\newc{\BR}{\hbox{\rm BR}}
\newc{\zbb}{Z\to b\bar}
\newc{\Gb}{\Gamma (Z\to b\bar b)}
\newc{\Gh}{\Gamma (Z\to \hbox{\rm hadrons})}
\newc{\rbsm}{R_b^\hbox{\rm sm}}
\newc{\rbsusy}{R_b^\hbox{\rm susy}}
\newc{\drb}{\delta R_b}
\newc{\sgn}{\mbox{sgn}}
\newc{\tbeta}{\tan\beta}
\newc{\uL}{{\tilde u_L}}
\newc{\uR}{{\tilde u_R}}
\newc{\cL}{{\tilde c_L}}
\newc{\cR}{{\tilde c_R}}
\newc{\tL}{{\tilde t_L}}
\newc{\tR}{{\tilde t_R}}
\newc{\dL}{{\tilde d_L}}
\newc{\dR}{{\tilde d_R}}
\newc{\sL}{{\tilde s_L}}
\newc{\sR}{{\tilde s_R}}
\newc{\bL}{{\tilde b_L}}
\newc{\bR}{{\tilde b_R}}
\newc{\eL}{{\tilde e_L}}
\newc{\eR}{{\tilde e_R}}
\newc{\mhp}{m_{H^\pm}}
\newc{\mhalf}{m_{1/2}}
\newc{\emt}{{e/\mu /\tau}}
\newc{\lR}{\tilde{l}_R}
\newc{\lL}{\tilde{l}_L}
\newc{\nL}{\tilde{\nu}_L}
\newc{\na}{\chi^0_1}
\newc{\nb}{\chi^0_2}
\newc{\ncc}{\chi^0_3}
\newc{\nd}{\chi^0_4}
\newc{\ca}{\chi^{\pm}_1}
\newc{\cb}{\chi^{\pm}_2}
\newc{\camp}{\chi^\mp_1}
\newc{\cbmp}{\chi^\mp_1}
\newc{\capos}{\chi^{+}_1}
\newc{\caneg}{\chi^{-}_1}
\newc{\phit}{\phi_t}
\newc{\phib}{\phi_b}
\newc{\phiew}{\phi_{ew}}
\newc{\htz}{h^0_t}
\newc{\hbz}{h^0_b}
\newc{\hewz}{h^0_{ew}}
\newc{\hsmz}{h^0_{sm}}
\newc{\huz}{h^0_u}
\newc{\hsusyz}{h^0_{susy}}

\def\beq{\begin{equation}}
\def\eeq{\end{equation}}
\def\bea{\begin{eqnarray}}
\def\eea{\end{eqnarray}}

\title{\Large\bf Particle Spectrum in the Minimal Supersymmetric Standard Model with non-universal Higgs masses}
\author{Levent  Solmaz\thanks{Department of Physics, Izmir Institute of Technology, IZTECH, Turkey
TR35430}\\ Department of Physics, Bal{\i}kesir University,
Bal{\i}kesir, Turkey, TR10100\\ E-mail:
\email{lsolmaz@balikesir.edu.tr}}

\abstract{We present semi-analytical solutions of the supersymmetric
non-universal masses models for  low $\tan\beta$ regime. In addition
to this, scale and $\tan\beta$ dependencies of the soft (mass)$^2$
terms are given in the form of numerical solutions. By using the
constrained form of the semi-analtic results, particular attention
is paid on the non-universality assumption of the Higgs mass values
and their potential measurable effects on the mass spectra of the
minimal supersymmetric standard model. It is observed that, certain
measurables are almost insensitive to the initial mass choices of
the Higgs fields, like the mass of the light $\mathcal{CP}$-even
Higgs boson. On the other hand, large deviations exist on the mass
of  the remaining physical Higgs bosons signal that the allowed
parameter space of the model can be probed successfully. For this
aim, in addition to the other physical Higgs bosons,  imprints
originating from the heavier chargino ($\tilde \chi^\pm_2$), heavy
neutralinos ($\tilde \chi^0_3$, $\tilde \chi^0_4$) and the light
scalar tau ($\tilde \tau_1$) are necessary and found to be
promising.}

\keywords{Supersymmetry Phenomenology }
\begin{document}
\section{Introduction}

There are a number of motivations for phenomenological studies of
the Supersymmetric (SUSY) theories among which  unification of the
gauge couplings and natural suppression of the radiative corrections
on the  masses of Higgs bosons can be mentioned (see {\it i.e.}
\cite{Chung:2003fi}, for a comprehensive list of motivations). Among
those theories, due to least number of particles, the Minimal
Supersymmetric Standard Model (MSSM) occupies a special place. In
the near future, forthcoming experiments may reveal that the
incorporation of the Standard Model (SM) into a more effective
theory turns out as the MSSM. Indeed, if low energy supersymmetry is
realized in Nature, phenomenological studies related with the MSSM
and its variants will be important to unreveal the hidden model.
Since it has certain problems like the famous $\mu$ problem
\cite{Kim:1983dt}, flavor problem \cite{Donoghue:1983mx}, and the
unknown mechanism of the supersymmetric symmetry breaking, studies
related with the extensions of the MSSM may be expected to  shed
light on future measurement, especially if nontrivial data
inconsistent with the minimal model occurs.

In this work, we  study particle spectrum in the MSSM with
non-universal Higgs mass terms (NUHM) \cite{Ellis:2006ix}. We
provide most general semi-analytic solutions of  evolving terms, in
terms of high scale boundary conditions, for a low $\tan\beta$
value. Additionally, different scale and $\tan\beta$ dependencies of
the soft (mass)$^2$ terms will be presented numerically. Actually,
the exploration of solutions to the renormalization group equations
(RGEs) of a supersymmetric model with NUHM is a subject that has
been investigated (see e.g. \cite{Huang:2000rn} and
\cite{Kazakov:1999pe}), but, the novel feature of our analysis is
that semi-analtic solutions may facilitate the exploration of the
phenomenology of the model (see \cite{Baer:2005bu} for phenomenology
of NUHM). As is well known, weak scale observables and Grand Unified
Theory (GUT) scale boundaries are connected via RGEs in a
complicated manner \cite{Martin:1993zk} and they  can be solved with
the help of certain softwares. Taste of numerical solutions can not
be compared with analytical ones though the former ones are very
accurate. As an alternative to the numerical ones, semi-analytic
expressions \cite{Kazakov:1999pe} and construction of certain RG
invariant forms are  useful for phenomenological analysis  of the
MSSM and its extensions \cite{ours}.

The possibility of non-universality specific to Higgs masses was
studied in a series of papers
\cite{Codoban:1999np},\cite{Ellis:2006ix}, by noting constraints
from $b\,\rightarrow\,s\gamma$, cosmology and anomalous magnetic
moment of muon and it was stressed that relaxing the scalar-mass
universality assumption for the MSSM Higgs multiplets opens up many
phenomenological possibilities (see also \cite{Ellis:2006jy} for
$B_s \rightarrow \mu^+\, \mu^-$ and cold dark matter issues related
with the NUHM). One of the aims of the present  work is to present
the full form of semi-analytical expressions explicitly, so that all
weak scale observables can be expressed in terms of GUT inputs. The
analytical form of the results can provide considerable insight for
similar issues (we ignore $\mathcal{CP}$-violation during the
numerical analysis of the NUHM, however, the full form of our
results cover this issue too). Indeed, due to the complicated
structure of the renormalization group equations
\cite{Martin:1993zk}, it is appealing to handle issues analytically
and the solutions presented in this work can be useful for such an
analysis even if they are given to the one loop order. As we will
see, to keep the analysis simple, there are certain ignorance made
on most of the correction terms, however, in the low $\tan\beta$
regime they do not affect our conclusions sizably and can further be
added on demand.

The outline of the rest of this work is as follows: In Section 2, we
introduce our notation and conventions. In Section 3 we present the
effects of non-universal Higgs masses terms on the supersymmetric
mass spectra for varying $\tan\beta$ and scale values. A subsection
of the same section is given to benchmark the semi-analytic results.
Section 4 is devoted to our conclusions. The full form of the
solutions of the RGEs can be found in the Appendix \ref{lowgen}.

\section{Notation and Conventions}

We define the basic parameters of the model as soft supersymmetry
breaking scalar masses $m_0$, gaugino masses $M$, the trilinear
couplings $A_0$, bilinear coupling $B_0$ and supersymmetric Higgs
mass parameter $\mu_0$, at the GUT scale. We assume third family
dominance model and solve RGEs at the one-loop order. In this
effective approach, by solving the RGEs explicitly, weak scale
predictions are expressed in terms of GUT boundaries. We express
Bino, Wino and Gluino with $M_{1,2,3}$, respectively, with a common
initial value $M$. By writing the GUT boundaries,
\begin{equation}
A_{i}=c_{A_i}\,A_0,\,\,\,M_j=c_{M_j}\,M,\,\,\,m_{k}=c_{k}\,m_0\,
\end{equation}
where $i=t,b,\tau$ and $j=1,2,3$, and for (mass)$^2$ terms
$k=H_u,H_d,{\tilde t_L},{\tilde t_R}, {\tilde b_R},{\tilde
\tau_L},{\tilde \tau_R}$. We will express weak scale value of each
quantity in terms of corresponding mSUGRA parameters $m_0,~M,~A_0$,
and a positive $\mu$ to be determined by the electroweak breaking
conditions. From the solutions of the RGEs, weak scale and GUT scale
values are connected and the most important restriction, in this
respect, is the mass of Z boson:
\begin{equation}
\label{MZconstraint}
\frac{1}{2}\,M^2_Z=-\mu^2 +\frac{m^2_{H_d}-\tan^2\beta\,m^2_{H_u}}{\tan^2\beta-1}+\Delta
\end{equation}
where $\tan\beta$ is the ratio of vacuum expectation values
($v_u/v_d$), and $\Delta$ stands for corrections on Higgs masses. We
are interested  in low vacuum expectation value ($\tan\beta=10$),
for which  complete list of semi-analytic solutions are given in the
Appendix \ref{lowgen}. In addition to this, we will present
graphical solutions for different $\tan\beta=10$ values in the next
section. Instead of purely numerical values, expressing weak scale
predictions in terms of GUT inputs proves very useful and helps to
differentiate the importance of each term. In order to show the
relative weigh of each term, we will express evolution of any soft
(mass)$^2$ term as in the following forms
\begin{equation}
\rm{(mass)}^2=\gamma_1\, A^2_0 + \gamma_2\, A_0\, M + \gamma_3\,
M^2  +  \gamma_4\, c^2_{H_d}\, m^2_0 +\gamma_5\, c^2_{H_u}\, m^2_0+ \gamma_6\, m^2_0\,.\\
  \label{nota}
  \end{equation}
 This decomposition enables one to lay stress upon the effects of
non-universal Higgs mass choices. As it can be extracted from the
above equation, sensitivity of each term to the initial values of
$m_{H_u}$ and $m_{H_d}$ will be different.  Notice that, by using
the $M_Z$ constraint given in (\ref{MZconstraint}) one can obtain
$\mu$ and this can be expressed as
\begin{eqnarray}
b=\frac{2\,|\mu|^2+m^2_{H_u}+m^2_{H_d}}{\tan\beta+\cot\beta}\,
\end{eqnarray}
hence, tree level relations of mass of physical Higgs boson can be
written as  in the followings \cite{Martin:1997ns}
\begin{eqnarray}
m^2_{A_0}&=&2\,b/\sin2\beta\\
m^2_{H^\pm}&=&m^2_{A_0}+m^2_W\\
m^2_{H_0,h_0}&=&\frac{1}{2}\left[m^2_{A_0}+m^2_{Z}\pm\sqrt{\left(m^2_{A_0}+m^2_{Z}\right)^2-4 m^2_{A_0}
m^2_{Z}\cos2\beta}~\right].
\end{eqnarray}
Those relations will be modified, largely, due to top-stop loop
corrections and $h_0$ is the most affected one. Indeed, since the
mass of the lightest $\mathcal{CP}$-even Higgs boson is larger than
$114\rm{\,GeV}$ \cite{LEPHiggs} this correction must be included in
the analysis. We will consider this correction and omit others in
our effective approach. Meanwhile, the price that should be paid for
that aim is predicting the spectra with small certain errors as will
be shown in the following section. But this does not affect our
conclusions, since  the reaction of the SUSY particles to the
non-universal Higgs boundary conditions is important for the present
study. The necessary expression for the most important correction is
\begin{equation}
\Delta(m^2_{h_0})=\frac{3}{4\,\pi^2}\,
v^2\,y^4_t\,\sin^4\beta\,\ln\left(\frac{m_{\tilde{t}_1}\,m_{\tilde{t}_2}}{m^2_t}\right)\,,
\end{equation}
where $m_{\tilde{t}_{1,2}}$ can be extracted from the following mass matrix
\begin{eqnarray}
\label{mzzp}
{\bold m^2_{\tilde t}}=\left(\begin{array}{cc}  {m}^2_t+m^2_{\tilde t_L}+(\frac{1}{2}-\frac{2}{3} s^2_w)\,M^2_Z \cos2\beta & {m}^2_t\left(A_t-\mu\cot\beta\right)\\\\
 {m}^2_t\left(A_t-\mu\cot\beta\right)& {m}^2_t+m^2_{\tilde t_R}+\frac{2}{3}\,s^2_w\,M^2_Z\,\cos2\beta\end{array}\right).
 \end{eqnarray}
 This $2\times2$ matrix can easily be diagonalized to obtain eigenvalues of stop quark
 masses in terms of GUT inputs,  similarly  the same should be done for ${\bold m^2_{\tilde b}}$, ${\bold
m^2_{\tilde \tau}}$ using the solutions presented in the appendices
in order to get the full sparticle spectrum as usual ({\it i.e.} see
\cite{Martin:1997ns}). Indeed, having such analytic expressions is
very useful to visualize the ingredients of sparticles to indirectly
probe the allowed range of non-universality of Higgs bosons. As an
example,  for  ($\tan\beta=10$)
 \begin{eqnarray}
\label{mtoptayfa} m^2_{{\tilde t}_{1,2}}&=&-0.0523\, A^2_0 + 0.192\, A_0\, M + 3.74\, M^2 + ( 0.642 -0.0176\, c^2_{H_d}- 0.161\, c^2_{H_u} ) \, m^2_0 +   {m}^2_t \nonumber\\
  &-& 0.245\, M^2_Z \mp \Omega
\end{eqnarray}
where the exact expression of $\Omega$ is a quite lengthy function
of all terms appearing in the first line of (\ref{mtoptayfa}).
Notice that, it can be obtained using the  full forms of the
solutions given in the Apppendix. Now, let us  make a simplifying
assumption $\mu_0\sim\,m_0\,\sim\,M\,\sim\,A_0$ and ${\bar
m_t}\sim\,2\,m_0$ on the $\Omega$ part of (\ref{mtoptayfa}) to
approximately predict the composition of  stop masses
\begin{eqnarray}
m^2_{\tilde t_{1,2}}&\simeq&-0.052\, A^2_0 + 0.19\, A_0\, M + 3.8\, M^2 + ( 0.64 -0.018\, c^2_{H_d}- 0.16\, c^2_{H_u} ) \, m^2_0 +   {m}^2_t \nonumber\\
  &-& 0.25 \, M^2_Z \mp\,3.45\,m^2_0~.\end{eqnarray}
Using this analytical expression, for instance, one can conclude
that weigh of up Higgs fields is larger than weigh of down Higgs
fields but their relative weigh is negligible compared to other soft
mass terms. To be specific, we will consider the specific reference
point $\rm{SPS1a^\prime}$ \cite{Aguilar-Saavedra:2005pw} in the
numerical analysis to benchmark the solutions provided. However,
even under the above rough approximation we found $m_{\tilde
t_1}=472\,\rm{\,GeV}$ and $m_{\tilde{t}_2}=506\,\rm{\,GeV}$, to be
compared with the exact results. For the mass spectra of SUSY
particles, effects of  up Higgs field can be dominant, however, as
we will see in the next section this can not be generalized to other
sectors.

\section{Numarical Analysis}
In this section $\tan\beta$ and scale evolutions of (mass)$^2$ terms
will be presented. For this aim, solutions of RGEs are performed
such that high scale is set equal to $1.9\times10^{16}\rm{\,GeV}$
and the supersymmetry breaking  scale is chosen as $1\rm{\,TeV}$.
With this choices, unification of the gauge couplings is satisfied
at the GUT scale as $g_1=g_2=g_3=0.718\pm0.001$.

One can obtain the solutions of RGEs for any $\tan\beta=10$. To be
specific, for $\tan\beta=10$, mass of the heavy SM fermions fix the
Yukawa couplings at the same scale as $Y_t=0.551,\, Y_b=0.0547,\,
Y_\tau=0.0685$. As a brief summary of the semi-analytic solutions,
this specific choice of $\tan\beta$ yields the followings equations
\begin{eqnarray}\label{lowspecial}
m^2_{H_u}&=&-0.102\, A^2_0 + 0.375\, A_0\, M -
1.93\, M^2 - 0.709\, m^2_0 +
  0.0331\, c^2_{H_d}\, m^2_0 +
  0.612\, c^2_{H_u}\, m^2_0\nonumber\\
m^2_{H_d}&=&-0.0107\, A^2_0 + 0.0309\, A_0\, M + 0.413\, M^2 -
0.0241\, m^2_0 +
  0.955\, c^2_{H_d}\, m^2_0 +
  0.0333\, c^2_{H_u}\, m^2_0\nonumber\\
m^2_{\tilde t_L}&=&-0.0367\, A^2_0 + 0.134\, A_0\, M + 4.33\, M^2 +
0.757\, m^2_0 +
  0.00768\, c^2_{H_d}\, m^2_0 -
  0.129\, c^2_{H_u}\, m^2_0\nonumber\\
m^2_{\tilde t_R}&=&-0.068\, A^2_0 + 0.25\, A_0\, M + 3.15\, M^2 +
0.527\, m^2_0 -
  0.0429\, c^2_{H_d}\, m^2_0 -
  0.194\, c^2_{H_u}\, m^2_0\,\,\\
m^2_{\tilde b_R}&=&-0.00534\, A^2_0 + 0.0192\, A_0\, M + 4.67\, M^2
+ 0.988\, m^2_0 +
  0.0149\, c^2_{H_d}\, m^2_0 -
  0.0211\, c^2_{H_u}\, m^2_0\nonumber\\
m^2_{\tilde \tau_L}&=&-0.00271\, A^2_0 + 0.00216\, A_0\, M + 0.493\,
M^2 + 0.994\, m^2_0 -
  0.0353\, c^2_{H_d}\, m^2_0 +
  0.0325\, c^2_{H_u}\, m^2_0\nonumber\\
m^2_{\tilde \tau_R}&=&-0.00542\, A^2_0 + 0.00432\, A_0\, M + 0.143\,
M^2 + 0.989\, m^2_0 +
  0.0595\, c^2_{H_d}\, m^2_0 -
  0.065\, c^2_{H_u}\, m^2_0\,.\nonumber
  \end{eqnarray}
Notice that the analytical expressions given in (\ref{lowspecial})
are constrained forms of the solutions presented in the Appendix
(here we set $\Phi_{i,j}\rightarrow\,0$ and $c_i\rightarrow\,1$,
except for $c_{H_u}$ and $c_{H_d}$). And they can be used at
$\rm{SPS1a^\prime}$ point \cite{Aguilar-Saavedra:2005pw}. We will
benchmark our solutions using this point in the following
subsection.

Different scale and $\tan\beta$ effects can be extracted from the
following figures (Figs. \ref{fig1}--\ref{fig7}). In Fig.
\ref{fig1}, we show $\tan\beta$ and scale dependencies of the
composition of $m^2_{H_u}$. Normally, mass of up Higgs fields get
contributions from any of the 28 terms given in (A.2). When we
assume CP is conserved ($\Phi_{i,j}\rightarrow\,0$) and accept
universality is in charge (except for Higgs fields), mass of the up
Higgs field can be decomposed in a neat form as in (2.3)
\begin{equation}
m^2_{H_u}=\gamma^{(H_u)}_1\, A^2_0 + \gamma^{(H_u)}_2\, A_0\, M +
\gamma^{(H_u)}_3\,
M^2  +  \gamma^{(H_u)}_4\, c^2_{(H_d)}\, m^2_0 +\gamma^{(H_u)}_5\, c^2_{(H_u)}\, m^2_0+ \gamma^{(H_u)}_6\, m^2_0\,.\\
  \label{nota}
  \end{equation}
As can be seen from both panels of the first figure, largest
contribution to mass of up Higgs field comes from Gaugino sector
(dashed-blue curves). Contribution of down Higgs field to up Higgs
field is negligible, in other words, deviation of down Higgs from
the universal choice can not  yield a detectable effect on up Higgs
field. In all Figs. \ref{fig1}--\ref{fig7}, solid red (green) curves
corresponds to contribution of $m^2_{H_d}$ ($m^2_{H_u}$) on the
related (mass)$^2$ terms, which are $m^2_{H_u}$,
$m^2_{H_d}$,$m^2_{t_L}$,$m^2_{t_R}$,$m^2_{b_R}$, $m^2_{l_L}$ and
finally $m^2_{l_R}$, respectively. In order to show the effects of
scale variations fix $\tan\beta=10$ (right panels) and for varying
$\tan\beta$ values scale is fixed around the weak scale (left
panels). The Figs. \ref{fig1}--\ref{fig7} denote that the
gauge/gaugino sector contributions to scalar mass sector evolution
increases scalar mass parameters as we go to the weak scale. It is
visible in Figs. \ref{fig6} and \ref{fig7} that a strong reaction
can be detected in the slepton sector to non-universal Higgs mass
terms and this is true for any $\tan\beta$ value. Notice that this
can be expected for Higgs bosons too (see Figs.
\ref{fig1}--\ref{fig2}). We observe from Figs.
\ref{fig3}--\ref{fig4} that, scalar top quarks are sensitive to NUHM
only for very small $\tan\beta$ values $(\sim\,2-3)$.
%%%%%%%%%%%%%%%%%%%%%%%%%%%%%%%%%%%%%%%%%%%%%%%%%%%%%%%%%%%%%%%%%%%

During the numerical analysis we observed that following the
physical Higgs bosons (except the CP-even light Higgs boson),
sleptons are very sensitive to NUHM terms. Hence, we present Fig.
\ref{fig8} to show a bird-eye picture of the reaction of stau mass
eigenvalues to NUHM parameters. As can be inferred from the very
figure, reaction of sparticles to the mentioned non-universality
drifts the mass predictions, to some extend. This effect ranges from
a few $\rm{GeV}$ to $\sim\,30-40\rm{\,GeV}$ for different sparticles
and it can be detectable since the correct spectrum is well known
for the MSSM. See Tab. \ref{table1} for the reaction of the
particles of the MSSM to NUHM.
  %%%%%%%%%%%%%%%%%%%%%%%%%%%%%%%%%%%%%%%%%%%%%%%%%%%%%%%%%%%%%%%%%%%
\begin{figure}[h!]
\begin{center}
\vspace{-2.9cm} \hspace{-.0cm}
\epsfig{file=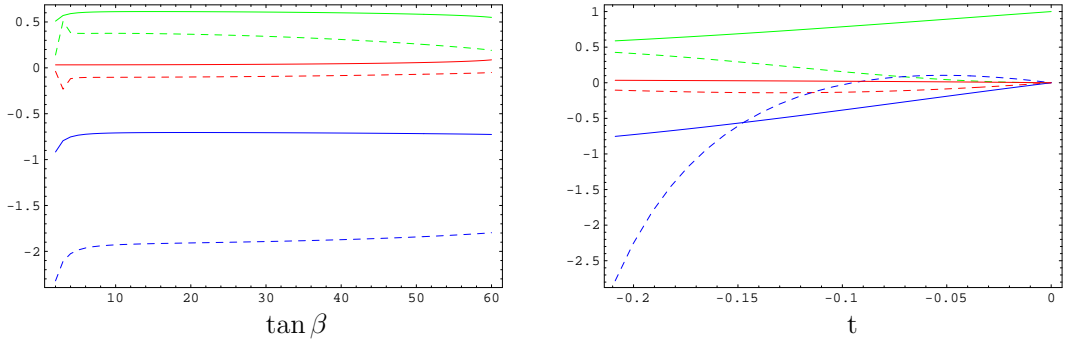,height=8.6in} %width=3.5in}\\
\vspace{-14.5cm}
\end{center}
\caption{Evolution of contributions for $m^2_{H_u}$ with $\tan\beta$
(left panel) and scale (right panel) in NUHM.  Scale is shown with
the dimensionless quantity t such that $\rm{t=0}$ denotes the GUT
scale and $\rm{t}\sim-0.2$ corresponds to the Z scale, $\tan\beta$
varies from 2 to 60. In both of the panels solid red, green and blue
lines correspond to $\gamma^{(H_u)}_4$, $\gamma^{(H_u)}_5$ and
$\gamma^{(H_u)}_6$. Dashed red, green and blue lines correspond to
$\gamma^{(H_u)}_1$, $\gamma^{(H_u)}_2$ and $\gamma^{(H_u)}_3$ as
given in (3.1)}\label{fig1}
\end{figure}
%%%%%%%%%%%%%%%%%%%%%%%%%%%%%%%%%%%%%%%%%%%%%%%%%%%%%%%%%%%%%%%%%%%
  %%%%%%%%%%%%%%%%%%%%%%%%%%%%%%%%%%%%%%%%%%%%%%%%%%%%%%%%%%%%%%%%%%%
\begin{figure}[h!]
\begin{center}
\vspace{-2.9cm} \hspace{-.0cm}
\epsfig{file=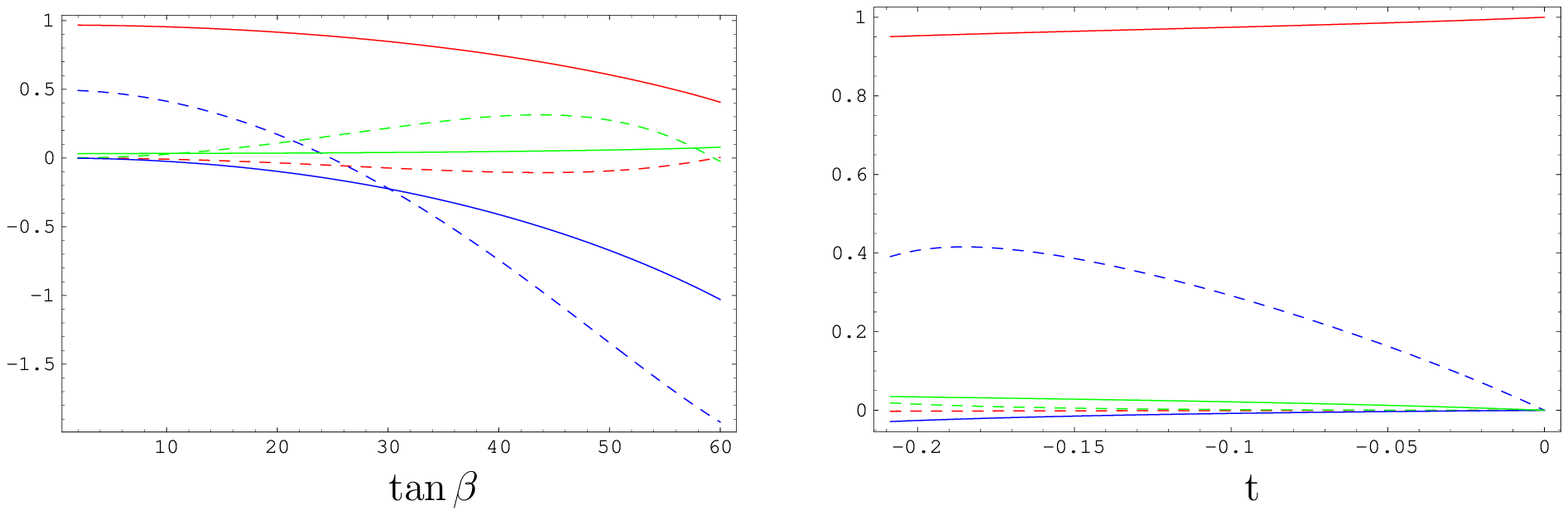,height=8.6in} %width=3.5in}\\
\vspace{-14.5cm}
\end{center}
\caption{The same as Fig. 1 but for $m^2_{H_d}$} \label{fig2}
\end{figure}
%%%%%%%%%%%%%%%%%%%%%%%%%%%%%%%%%%%%%%%%%%%%%%%%%%%%%%%%%%%%%%%%%%%
 %%%%%%%%%%%%%%%%%%%%%%%%%%%%%%%%%%%%%%%%%%%%%%%%%%%%%%%%%%%%%%%%%%%
\begin{figure}[h!]
\begin{center}
\vspace{-2.9cm} \hspace{0cm}
\epsfig{file=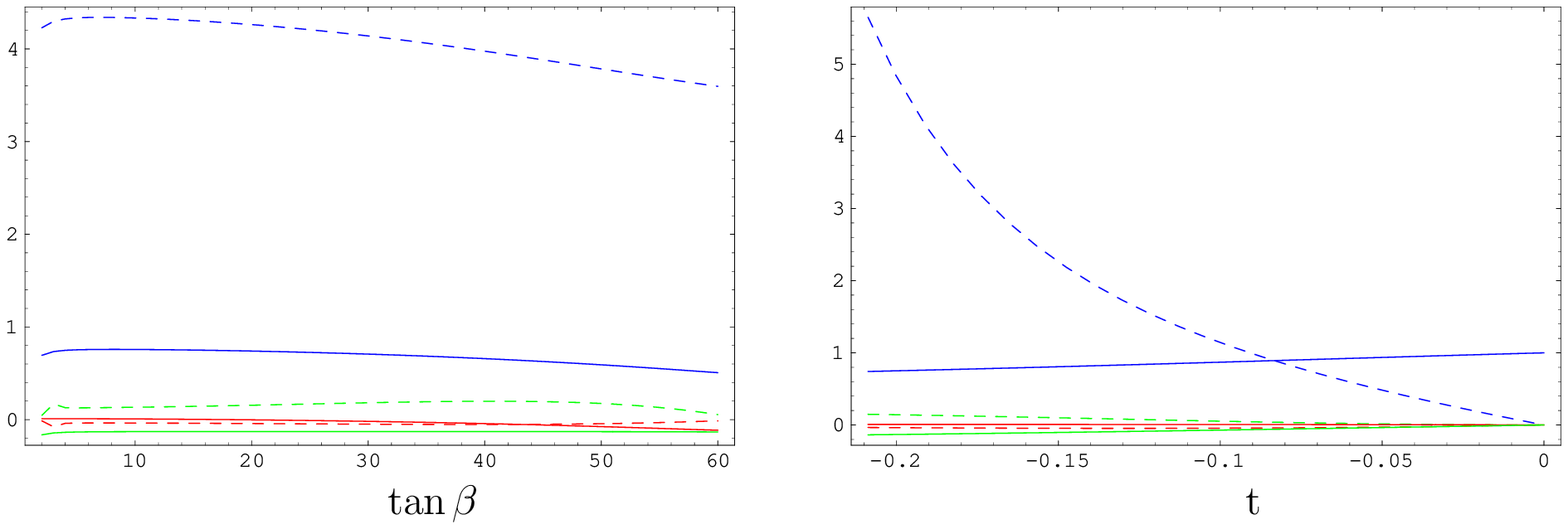,height=8.6in} %width=3.5in}\\
\vspace{-14.5cm}
\end{center}
\caption{The same as Fig. 1 but for $m^2_{t_L}$} \label{fig3}
\end{figure}
%%%%%%%%%%%%%%%%%%%%%%%%%%%%%%%%%%%%%%%%%%%%%%%%%%%%%%%%%%%%%%%%%%%
 %%%%%%%%%%%%%%%%%%%%%%%%%%%%%%%%%%%%%%%%%%%%%%%%%%%%%%%%%%%%%%%%%%%
\begin{figure}[h!]
\begin{center}
\vspace{-2.9cm} \hspace{0cm}
\epsfig{file=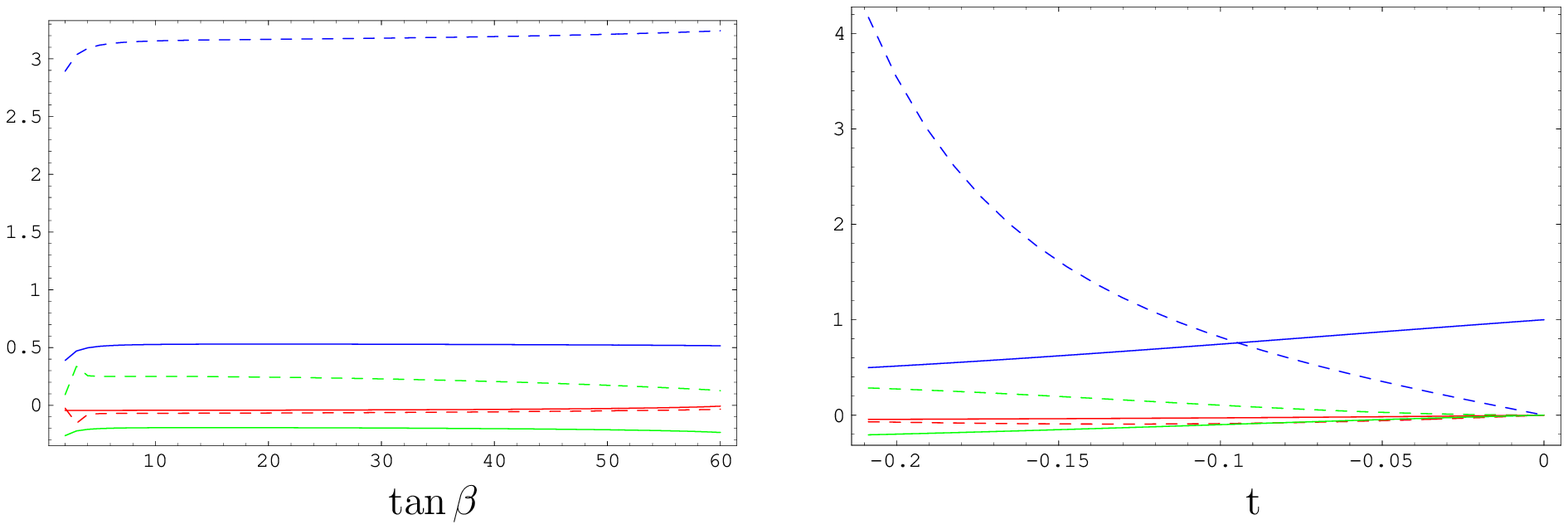,height=8.6in} %width=3.5in}\\
\vspace{-14.5cm}
\end{center}
\caption{The same as Fig. 1 but for $m^2_{t_R}$} \label{fig4}
\end{figure}
%%%%%%%%%%%%%%%%%%%%%%%%%%%%%%%%%%%%%%%%%%%%%%%%%%%%%%%%%%%%%%%%%%%
 %%%%%%%%%%%%%%%%%%%%%%%%%%%%%%%%%%%%%%%%%%%%%%%%%%%%%%%%%%%%%%%%%%%
\begin{figure}[h!]
\begin{center}
\vspace{-2.9cm} \hspace{0cm}
\epsfig{file=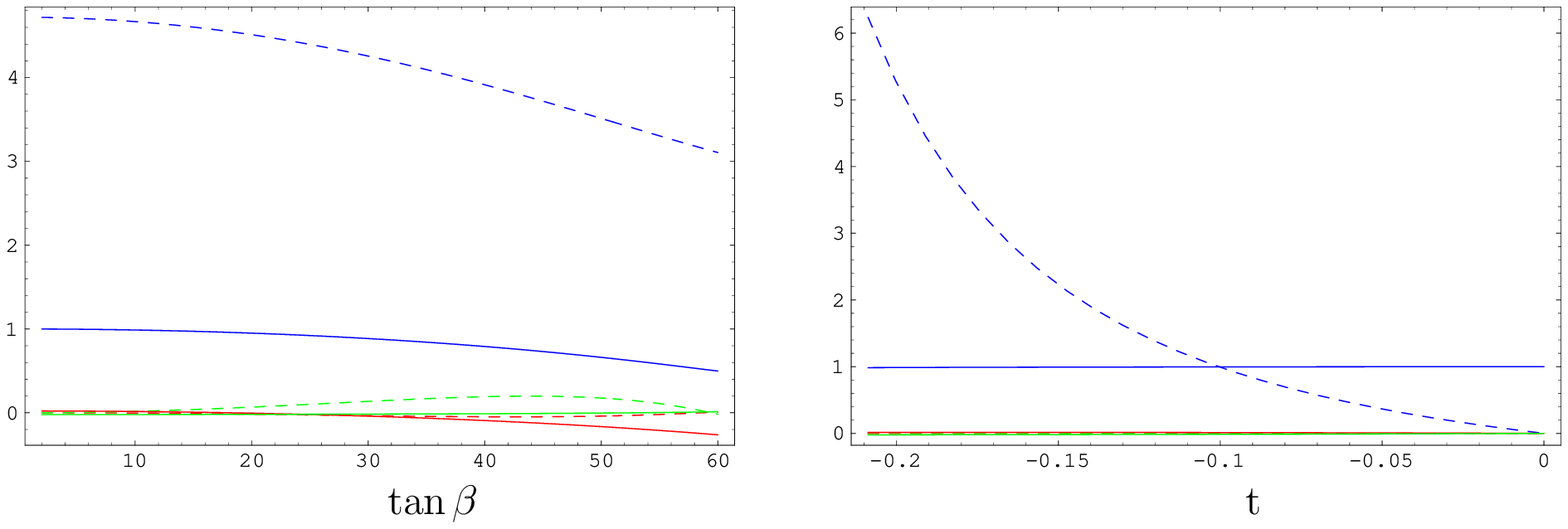,height=8.6in} %width=3.5in}\\
\vspace{-14.5cm}
\end{center}
\caption{The same as Fig. 1 but for $m^2_{b_R}$} \label{fig5}
\end{figure}
%%%%%%%%%%%%%%%%%%%%%%%%%%%%%%%%%%%%%%%%%%%%%%%%%%%%%%%%%%%%%%%%%%%
%%%%%%%%%%%%%%%%%%%%%%%%%%%%%%%%%%%%%%%%%%%%%%%%%%%%%%%%%%%%%%%%%%%
\begin{figure}[h!]
\begin{center}
\vspace{-2.9cm} \hspace{0cm}
\epsfig{file=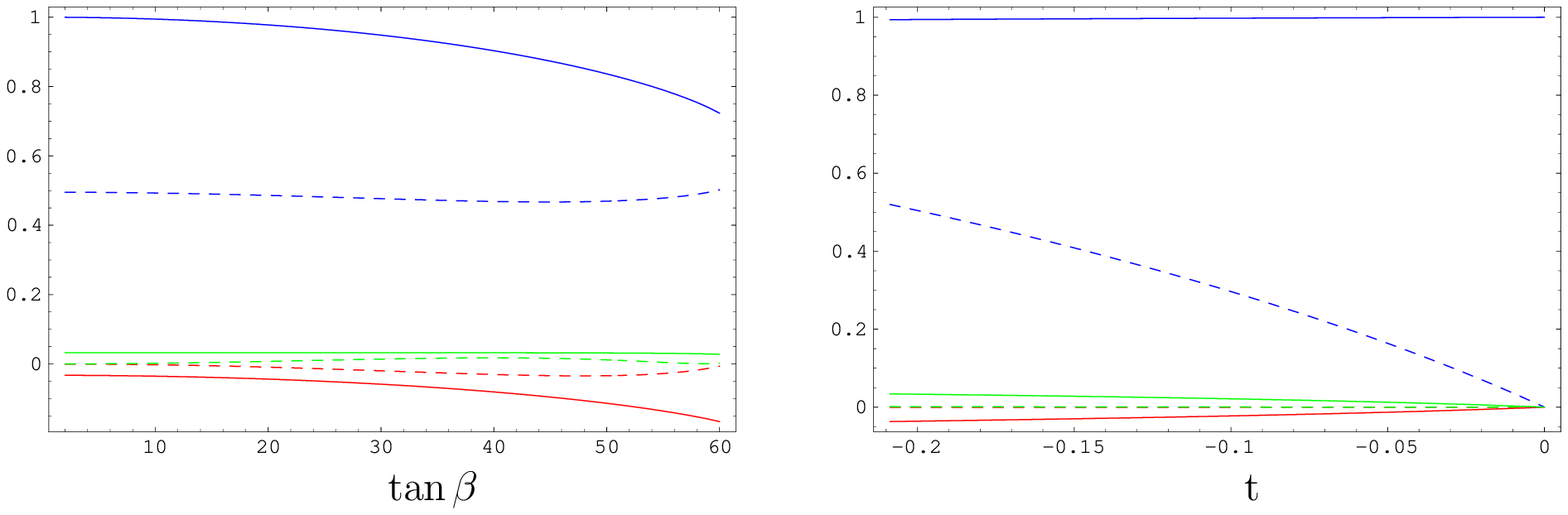,height=8.6in} %width=3.5in}\\
\vspace{-14.5cm}
\end{center}
\caption{The same as Fig. 1 but for $m^2_{l_L}$} \label{fig6}
\end{figure}
%%%%%%%%%%%%%%%%%%%%%%%%%%%%%%%%%%%%%%%%%%%%%%%%%%%%%%%%%%%%%%%%%%%
 %%%%%%%%%%%%%%%%%%%%%%%%%%%%%%%%%%%%%%%%%%%%%%%%%%%%%%%%%%%%%%%%%%%
\begin{figure}[h!]
\begin{center}
\vspace{-2.9cm} \hspace{0cm}
\epsfig{file=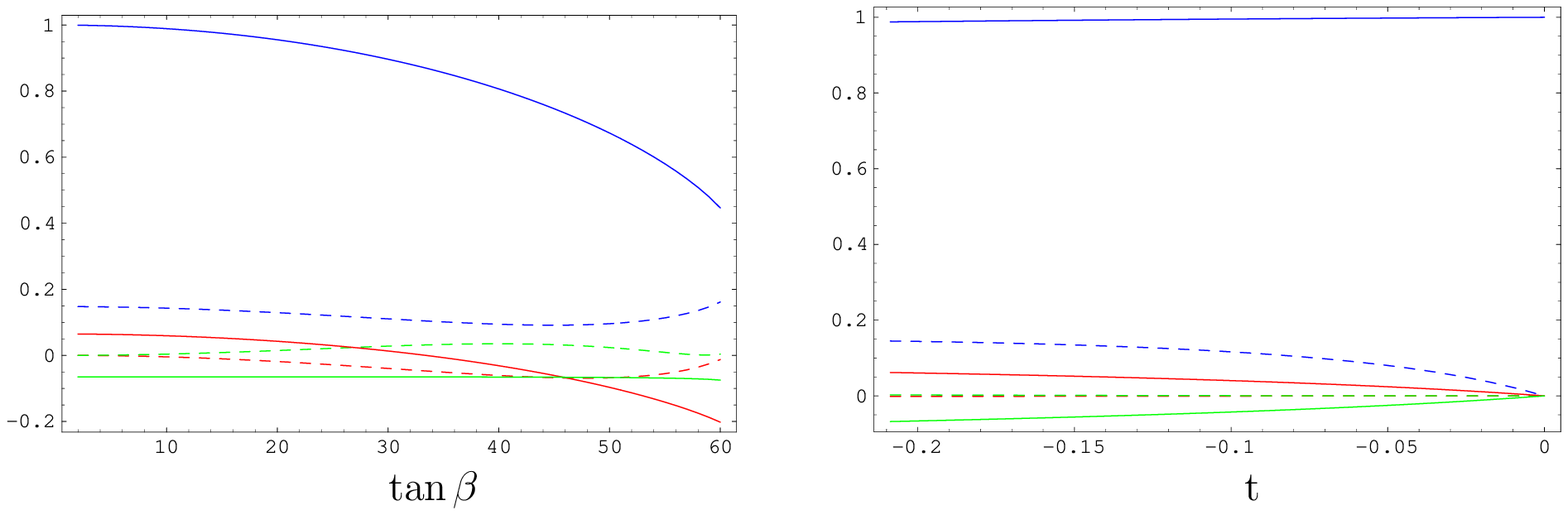,height=8.6in} %width=3.5in}\\
\vspace{-14.5cm}
\end{center}
\caption{The same as Fig. 1 but for $m^2_{l_R}$} \label{fig7}
\end{figure}
%%%%%%%%%%%%%%%%%%%%%%%%%%%%%%%%%%%%%%%%%%%%%%%%%%%%%%%%%%%%%%%%%%%
%%%%%%%%%%%%%%%%%%%%%%%%%%%%%%%%%%%%%%%%%%%%%%%%%%%%%%%%%%%%%%%%%%%
\begin{figure}[t!]
\begin{center}
\vspace{-3.cm} \hspace{-.0cm}
\epsfig{file=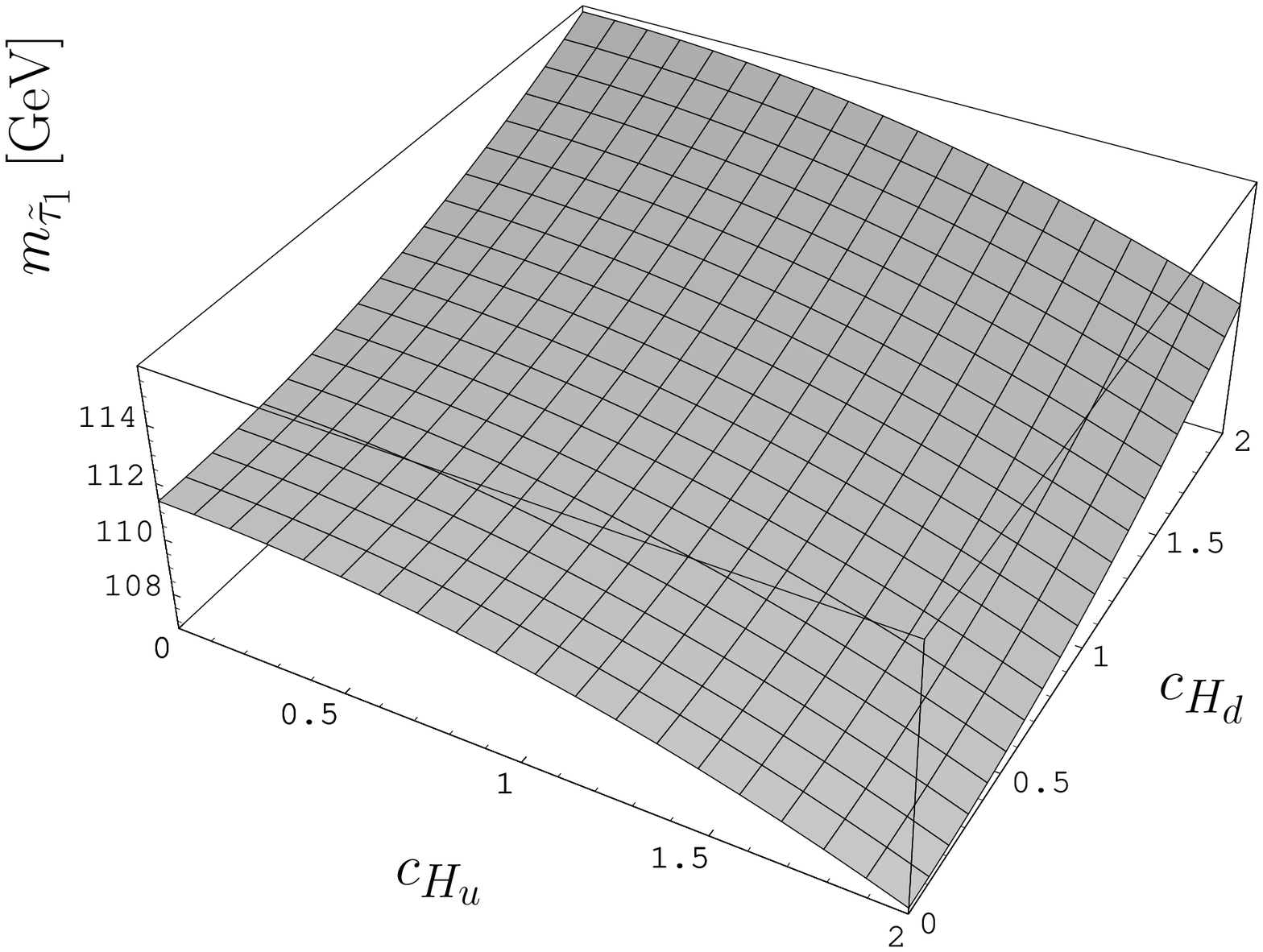,height=4in} %width=3.5in}
\epsfig{file=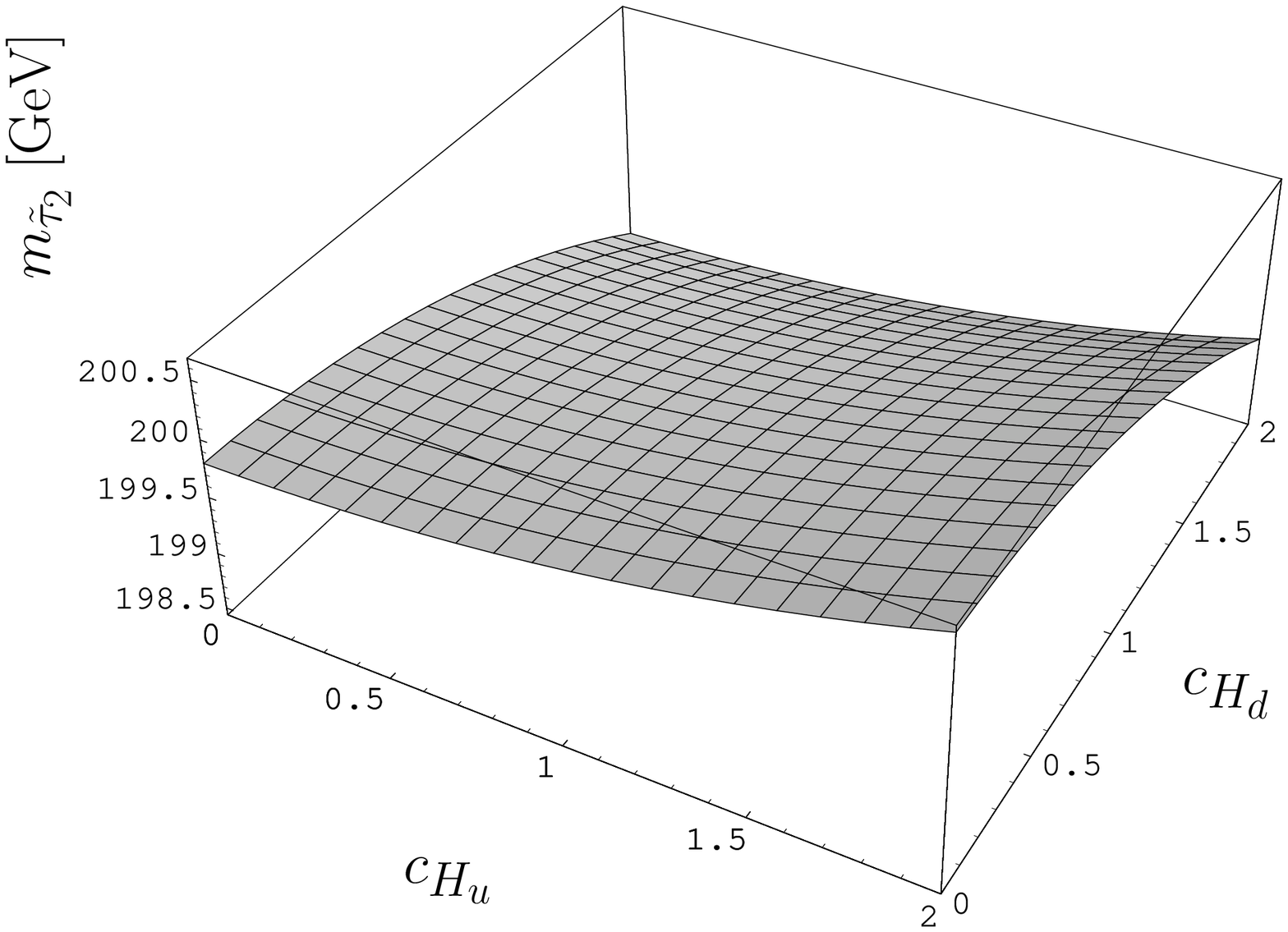,height=4in} %width=3.5in}
\vspace{-2.2cm}
\end{center}
\caption{3D plot of variation stau mass eigenvalues, with
non-universal coefficients at the $\rm{SPS1a^\prime}$ point. The
case $c_{H_u}=c_{H_d}=1$
  brings $m_{\tilde \tau_1}=\rm{\,111.4\,GeV}$ (left) and $m_{\tilde \tau_2}=199.7\rm{\,GeV}$ (right).
  Stop sensitivities are $\delta^{\rm{NUHM}}_{\tilde \tau_1}\sim8.8\rm{\,GeV}$ and
   $\delta^{\rm{NUHM}}_{\tilde \tau_2}\sim2.1\rm{\,GeV}$.}
\label{fig8}
\end{figure}
%%%%%%%%%%%%%%%%%%%%%%%%%%%%%%%%%%%%%%%%%%%%%%%%%%%%%%%%%%%%%%%%%%%

\subsection{Benchmark of the solutions}
The most practical solution in order to test the trustability of our
results is to use certain benchmark points. Though a large set of
benchmark points and parameter lines in the MSSM parameter space is
established, we will use one the the most studied points (see
\cite{Allanach:2002nj} for Snowmass Points and Slopes). Since we
ignored  most of the corrections except that of on the mass of the
lightest $\mathcal{CP}$-even Higgs boson, a strict comparison with
the state of art programs like
 ISAJET \cite{isajet} or SOFTSUSY \cite{softsusy}  shoul not be expected (see also \cite{kram}
 and the given web page for online comparison). Nevertheless, resulting error  should not be too high and
there should be a visible correlation. We observed this is indeed
the case for our semi-analytic solutions. To be definite, if
$\tan\beta=10$, $M=250\rm{\,GeV}$, $m_0=70\rm{\,GeV}$,
$A_0=-300\rm{\,GeV}$, $\frac{\mu}{|\mu|}=1$ (which is the
$\rm{SPS1a^\prime}$ reference point) then at the weak scale (at 1
$\rm{\,TeV}$) we end up with Table \ref{table1}.
%------------------------------------------------------------------------------------
\begin{table}
\begin{center}
\begin{tabular}{c c c c ||c}
  Particle & $SPS1a^\prime$ [GeV] & This Work [GeV] &  $\rm{\%}$ Difference  &  $\delta^{\rm{NUHM}}$ [GeV]\\
 \hline\hline
$h^0$   & 116.0     &  110.3&   4.91        &   0.2   \\
$H^0$   & 425.0     &  425.7&   -0.17       &   35.1  \\
$A^0$   & 424.9     &  425.3&   -0.09       &   35.1  \\
$H^{\pm}$         & 432.7     &  432.8&   -0.02       &   34.5  \\
$\tilde{t}_1$     & 366.5     &  374.3&   -2.13       &   5.1   \\
$\tilde{t}_2$     & 585.5     &  578.9&   1.13        &   5.1   \\
$\tilde{b}_1$     & 506.3     &  502.9&   0.67        &   2.4   \\
$\tilde{b}_2$     & 545.7     &  530.7&   2.75        &   0.9   \\
$\tilde{\tau}_1$  & 107.9     &  111.4&   -3.24       &   8.8   \\
$\tilde{\tau}_2$  & 194.9     &  199.7&   -2.46       &   2.1  \\
$\tilde{\chi}^0_{1}$        &  97.7     &  105.3&   -7.78       &   0.2   \\
$\tilde{\chi}^0_{2}$        & 183.9     &  194.3&   -5.65       &   1.2   \\
$\tilde{\chi}^0_{3}$        & 400.5     &  400.6&   -0.02       &   15.9  \\
$\tilde{\chi}^0_{4}$        & 413.9     &  417.3&   -0.82       &   14.5  \\
$\tilde{\chi}^{\pm}_{1}$    & 183.7     &  193.7&   -5.44       &   1.3  \\
$\tilde{\chi}^{\pm}_{2}$    & 415.4     &  417.9&   -0.60       &   14.6  \\
$\tilde{\upsilon}_{\tau}$   & 170.5     &  176.6&   -3.58       &   3.8   \\
\hline\hline
\end{tabular}
\end{center}
\caption{\label{table1}}{Numerical values for the mass of some of
the supersymmetric particles and Higgs bosons in the reference point
$\rm{SPS1a^\prime}$ \cite{Aguilar-Saavedra:2005pw} and their
comparison with our semi-analytical results. The third column is
obtained by $(SPS1a^\prime-\rm{our~results})\times
100/SPS1a^\prime$. The fourth column denotes the sensitivity of each
particle to the NUHM model parameters. The difference between
maximal and minimal mass values is obtained by varying $c_{H_u}$ and
$c_{H_d}$ in the $[0,2]$ interval and the emerging difference is
called as sensitivity ($\delta^{\rm{NUHM}}$) for each term.}
\end{table}

 Comparison of these results with the reference point denotes that the errors in predicting $m_h$, $m_{{\tilde \tau}_{1,2}}$ and $m_{\widetilde{\chi}^0_{1}}$
are negligible. For other mass terms errors are  somewhat large
especially in predicting mass of the lightest neutralino
$m_{\tilde{\chi}^0_1}$; here absolute error is $\sim 8 \rm{\%}$
which could be reduced if calculation are performed at two loops,
corrections are noticed for all  terms etc. However, apparent
correlation is sufficient for our aim since we are basically
interested in the reaction of those particles  to the non-universal
choices of the Higgs masses. Of course, this is true as far as
corrections do not alter the weight of $c_{H_u}$ and $c_{H_d}$ on
the SUSY particles, which we assumed to be true since the emerging
mass difference of the worst prediction is less than $\sim 8
\rm{\%}$. Nevertheless, a numerical simulation including all
families and known corrections would be more decisive, which is
beyond the scope of this work.

\section{Conclusions}
Using the semi-analytic solutions presented in this work it is
observed that  deviation from the universality assumption of the
Higgs fields does not induce serious problems as in the case of
other soft $(\rm{mass})^2$ terms (especially  if  $c_{H_u}\sim
c_{H_d}\sim1$). This can be inferred from Tab \ref{table1} in which
coefficients of up and down Higgs fields are varied from 0 to 2
$m_0$.  For this range, a striking difference can be observed on the
mass of certain supersymmetric particles (like sleptons) while the
others are insensitive to the mentioned phenomena (like lightest
neutralino).

Expected discovery of low energy SUSY  at the Large Hadron Collider
(LHC) and the International Linear Collider (ILC)
\cite{Weiglein:2004hn} will require reconstruction of the
supersymmetric theory parameters from the experimental data. This is
necessary  not only for the minimal model but also for NUHM,
especially if experimental data signalling deviations from the
minimal supergravity model (mSUGRA) \cite{mSUGRA} occurs. For this
aim, precise measurements of mass of the light stau $m_{\tau_1}$,
which is probably among the first sparticles to be discovered due to
lepton nature and a light mass very sensitive to non-universality of
the Higgs bosons, will be very suitable to shoot the NUHM parameter
space. \acknowledgments I m thankful to Durmus A. Demir for reading
the manuscript and useful discussions. This work is partially
supported by post-doctoral fellowship of the Scientific and
Technical Research Council of Turkey.

\appendix
\label{APP}
\section{Explicit Solutions for \bf{low} $\tan\beta$}
\label{lowgen} In this part we present explicit form of our
semi-analytic solutions which are obtained by solving the RGEs
explicity, to the one loop order. The Gut scale is $M_{GUT}=1.9
\times 10^{16}\rm{\,GeV}$ and
 $\tan\beta=10$. At the GUT scale we found the following results for gauge and Yukawa couplings
\begin{eqnarray}
& & g_1=0.7179,\,g_2=0.7187,\, g_3=0.7195\,\nonumber\\
& & Y_t=0.5510,\, Y_b=0.0547,\, Y_\tau=0.0685.\,\nonumber\\
\end{eqnarray}
For the weak scale scale ($\sim\,1\rm{~TeV}$), our results for soft
$(\rm{mass})^2$ terms read
\begin{eqnarray}
m^2_{H_u}&=&0.000619\, A^2_0\, c^2_{A_b} -
  7.8\, \times\, {10}^{-7}\, A^2_0\, c^2_{A_\tau} -
  0.103\, A^2_0\, c^2_{A_t} +
  0.00473\, c^2_{M_1}\, M^2 \,\nonumber\\
  &+& 0.206\, c^2_{M_2}\,
M^2 - 1.94\, c^2_{M_3}\, M^2 + 0.0331\, c^2_{H_d}\, m^2_0 +
  0.612\, c^2_{H_u}\, m^2_0 -
  0.0319\, c^2_{b_R}\, m^2_0 \,\nonumber\\
  &+&
  0.0325\, c^2_{\tau_L}\, m^2_0 -
  0.0325\, c^2_{\tau_R}\, m^2_0 -
  0.387\, c^2_{t_L}\, m^2_0 -
  0.29\, c^2_{t_R}\, m^2_0 \,\nonumber\\
  &-& 0.00572\, c_{M_1}\, c_{M_2}\,M^2\, \cos \Phi_{12} - 0.0252\, c_{M_1}\, c_{M_3}\, M^2\, \cos \Phi_{13} \,\nonumber\\
  &-& 0.0000612\, A_0\, c_{A_b}\, c_{M_1}\, M\, \cos \Phi_{1b} +  2.13\, \times\, {10}^{-7}\, A_0\, c_{A_\tau}\, c_{M_1}\,M\, \cos \Phi_{1\tau} \,\nonumber\\
  &+& 0.0122\, A_0\, c_{A_t}\, c_{M_1}\, M\, \cos\Phi_{1t}- 0.168\, c_{M_2}\, c_{M_3}\, M^2\, \cos \Phi_{23} \,\nonumber\\
  &-& 0.000535\, A_0\, c_{A_b}\, c_{M_2}\, M\, \cos\Phi_{2b} + 1.12\, \times\, {10}^{-6}\, A_0\, c_{A_\tau}\, c_{M_2}\, M\, \cos \Phi_{2\tau} \,\nonumber\\
  &+& 0.0726\, A_0\, c_{A_t}\, c_{M_2}\, M\, \cos\Phi_{2t} - 0.00215\, A_0\, c_{A_b}\, c_{M_3}\, M\, \cos \Phi_{3b} \,\nonumber\\
  &+& 3.48\, \times\, {10}^{-6}\, A_0\, c_{A_\tau}\, c_{M_3}\, M\, \cos \Phi_{3\tau} +  0.293\, A_0\, c_{A_t}\, c_{M_3}\, M\, \cos\Phi_{3t} \,\nonumber\\
  &-& 1.54\, \times\, {10}^{-6}\, A^2_0\, c_{A_b}\, c_{A_\tau}\, \cos \Phi_{b\tau} + 0.000285\, A^2_0\, c_{A_b}\, c_{A_t}\, \cos \Phi_{tb} \,\nonumber\\
  &-& 3.29\, \times\, {10}^{-7}\, A^2_0\, c_{A_\tau}\, c_{A_t}\, \cos \Phi_{t\tau}~,\,\end{eqnarray}
\begin{eqnarray}
 m^2_{H_d}&=&-0.00992\, A^2_0\, c^2_{A_b} -
    0.00272\, A^2_0\, c^2_{A_\tau} +
    0.000286\, A^2_0\, c^2_{A_t} +
    0.0361\, c^2_{M_1}\, M^2 \,\nonumber\\
  &+& 0.449\, c^2_{M_2}\,
  M^2 - 0.0613\, c^2_{M_3}\,
  M^2 + 0.955\, c^2_{H_d}\, m^2_0 +
    0.0333\, c^2_{H_u}\, m^2_0 +
    0.0224\, c^2_{b_R}\, m^2_0 \,\nonumber\\
  &-&    0.0353\, c^2_{\tau_L}\, m^2_0 +
    0.0298\, c^2_{\tau_R}\, m^2_0 +
    0.0232\, c^2_{t_L}\, m^2_0 -
    0.0642\, c^2_{t_R}\, m^2_0 \,\nonumber\\
  &-& 0.000383\, c_{M_1}\, c_{M_2}\,   M^2\, \cos \Phi_{12} -     0.000749\, c_{M_1}\, c_{M_3}\,   M^2\, \cos \Phi_{13} \,\nonumber\\
  &+& 0.000538\, A_0\, c_{A_b}\, c_{M_1}\,   M\, \cos \Phi_{1b} +     0.000586\, A_0\, c_{A_\tau}\, c_{M_1}\,  M\, \cos \Phi_{1\tau} \,\nonumber\\
  &-& 0.0000797\, A_0\, c_{A_t}\, c_{M_1}\,  M\, \cos \Phi_{1t} - 0.0097\, c_{M_2}\, c_{M_3}\,   M^2\, \cos \Phi_{23} \,\nonumber\\
  &+& 0.0064\, A_0\, c_{A_b}\, c_{M_2}\,   M\, \cos \Phi_{2b} +     0.0016\, A_0\, c_{A_\tau}\, c_{M_2}\,  M\, \cos \Phi_{2\tau} \,\nonumber\\
  &-& 0.000762\, A_0\, c_{A_t}\, c_{M_2}\,   M\, \cos \Phi_{2t} +    0.0258\, A_0\, c_{A_b}\, c_{M_3}\,  M\, \cos \Phi_{3b} \,\nonumber\\
  &-& 0.0000721\, A_0\, c_{A_\tau}\, c_{M_3}\,   M\, \cos \Phi_{3\tau} -    0.00307\, A_0\, c_{A_t}\, c_{M_3}\,  M\, \cos \Phi_{3t} \,\nonumber\\
  &+& 0.0000554\, A^2_0\, c_{A_b}\, c_{A_\tau}\, \cos \Phi_{b\tau} +     0.00159\, A^2_0\, c_{A_b}\, c_{A_t}\, \cos \Phi_{tb} \,\nonumber\\
  &-& 4.46\, \times\, {10}^{-6}\, A^2_0\, c_{A_\tau}\, c_{A_t}\, \cos \Phi_{t\tau}~,\,\end{eqnarray}
\begin{eqnarray}
m^2_{\tilde t_L}&=&-0.0031\, A^2_0\, c^2_{A_b} +
    5.35\, \times\, {10}^{-6}\, A^2_0\, c^2_{A_\tau} -
    0.0342\, A^2_0\, c^2_{A_t} -
    0.00678\, c^2_{M_1}\, M^2 \,\nonumber\\
  &+& 0.372\, c^2_{M_2}\,
  M^2 + 4.04\, c^2_{M_3}\,
  M^2 + 0.00768\, c^2_{H_d}\, m^2_0 -
    0.129\, c^2_{H_u}\, m^2_0 -
    0.014\, c^2_{b_R}\, m^2_0 \,\nonumber\\
  &+&   0.0108\, c^2_{\tau_L}\, m^2_0 -
    0.0108\, c^2_{\tau_R}\, m^2_0 +
    0.868\, c^2_{t_L}\, m^2_0 -
    0.0965\, c^2_{t_R}\, m^2_0 \,\nonumber\\
  &-&    0.00197\, c_{M_1}\, c_{M_2}\,   M^2\, \cos \Phi_{12} -     0.00865\, c_{M_1}\, c_{M_3}\,   M^2\, \cos \Phi_{13} \,\nonumber\\
  &+&     0.00016\, A_0\, c_{A_b}\, c_{M_1}\,   M\, \cos \Phi_{1b} -     1.29\,       \times\, {10}^{-6}\, A_0\, c_{A_\tau}\, c_{M_1}\, M\, \cos \Phi_{1\tau} \,\nonumber\\
  &+&    0.00403\, A_0\, c_{A_t}\, c_{M_1}\,   M\, \cos \Phi_{1t} - 0.0592\, c_{M_2}\, c_{M_3}\,   M^2\, \cos \Phi_{23} \,\nonumber\\
  &+&    0.00196\, A_0\, c_{A_b}\, c_{M_2}\,   M\, \cos \Phi_{2b} -     6.44\,      \times\, {10}^{-6}\, A_0\, c_{A_\tau}\, c_{M_2}\, M\, \cos \Phi_{2\tau} \,\nonumber\\
  &+&    0.0239\, A_0\, c_{A_t}\, c_{M_2}\,  M\, \cos \Phi_{2t} +    0.00789\, A_0\, c_{A_b}\, c_{M_3}\,  M\, \cos \Phi_{3b} \,\nonumber\\
  &-&    0.000017\, A_0\, c_{A_\tau}\, c_{M_3}\,  M\, \cos \Phi_{3\tau} +    0.0965\, A_0\, c_{A_t}\, c_{M_3}\,  M\, \cos \Phi_{3t} \,\nonumber\\
  &+&    0.0000106\, A^2_0\, c_{A_b}\, c_{A_\tau}\, \cos\Phi_{b\tau} +    0.000627\, A^2_0\, c_{A_b}\, c_{A_t}\, \cos\Phi_{tb} \,\nonumber\\
  &-&    1.17\,      \times\, {10}^{-6}\, A^2_0\, c_{A_\tau}\,c_{A_t}\, \cos \Phi_{t\tau}~,\,\end{eqnarray}
\begin{eqnarray}
m^2_{\tilde t_R}&=&0.000412\, A^2_0\, c^2_{A_b} -
  5.2\, \times\, {10}^{-7}\, A^2_0\, c^2_{A_\tau} -
  0.0686\, A^2_0\, c^2_{A_t} +
  0.0443\, c^2_{M_1}\, M^2 \,\nonumber\\
  &-& 0.168\, c^2_{M_2}\,
M^2 + 3.41\, c^2_{M_3}\, M^2 - 0.0429\, c^2_{H_d}\, m^2_0 -
  0.194\, c^2_{H_u}\, m^2_0 +
  0.0438\, c^2_{b_R}\, m^2_0 \,\nonumber\\
  &-&
  0.0434\, c^2_{\tau_L}\, m^2_0 +
  0.0434\, c^2_{\tau_R}\, m^2_0 -
  0.193\, c^2_{t_L}\, m^2_0 +
  0.676\, c^2_{t_R}\, m^2_0 \,\nonumber\\
  &-&   0.00381\, c_{M_1}\, c_{M_2}\,M^2\, \cos \Phi_{12} - 0.0168\, c_{M_1}\, c_{M_3}\, M^2\, \cos \Phi_{13} \,\nonumber\\
  &-&  0.0000408\, A_0\, c_{A_b}\, c_{M_1}\, M\, \cos \Phi_{1b} +  1.42\,    \times\, {10}^{-7}\, A_0\, c_{A_\tau}\, c_{M_1}\,M\, \cos \Phi_{1\tau} \,\nonumber\\
  &+&  0.0081\, A_0\, c_{A_t}\, c_{M_1}\, M\, \cos \Phi_{1t} - 0.112\, c_{M_2}\, c_{M_3}\, M^2\, \cos \Phi_{23} \,\nonumber\\
  &-&  0.000357\, A_0\, c_{A_b}\, c_{M_2}\, M\, \cos\Phi_{2b} +  7.45\,    \times\, {10}^{-7}\, A_0\, c_{A_\tau}\, c_{M_2}\,M\, \cos \Phi_{2\tau} \,\nonumber\\
  &+&  0.0484\, A_0\, c_{A_t}\, c_{M_2}\, M\, \cos\Phi_{2t} -  0.00144\, A_0\, c_{A_b}\, c_{M_3}\, M\, \cos\Phi_{3b} \,\nonumber\\
  &+&  2.32\,    \times\, {10}^{-6}\, A_0\, c_{A_\tau}\, c_{M_3}\,M\, \cos \Phi_{3\tau} +  0.195\, A_0\, c_{A_t}\, c_{M_3}\, M\, \cos\Phi_{3t} \,\nonumber\\
  &-&  1.03\, \times\, {10}^{-6}\, A^2_0\, c_{A_b}\,c_{A_\tau}\, \cos \Phi_{b\tau} +  0.00019\, A^2_0\, c_{A_b}\, c_{A_t}\, \cos\Phi_{tb} \,\nonumber\\
  &-&  2.19\,    \times\, {10}^{-7}\, A^2_0\, c_{A_\tau}\,c_{A_t}\, \cos \Phi_{t\tau}~,\,\end{eqnarray}
\begin{eqnarray}
m^2_{\tilde b_R}&=&-0.00662\, A^2_0\, c^2_{A_b} +
  0.0000112\, A^2_0\, c^2_{A_\tau} +
  0.000191\, A^2_0\, c^2_{A_t} +
  0.0162\, c^2_{M_1}\, M^2 \,\nonumber\\
  &-& 0.00483\, c^2_{M_2}\,
M^2 + 4.66\, c^2_{M_3}\, M^2 + 0.0149\, c^2_{H_d}\, m^2_0 -
  0.0211\, c^2_{H_u}\, m^2_0 +
  0.972\, c^2_{b_R}\, m^2_0 \,\nonumber\\
  &+&
  0.0217\, c^2_{\tau_L}\, m^2_0 -
  0.0217\, c^2_{\tau_R}\, m^2_0 -
  0.0279\, c^2_{t_L}\, m^2_0 +
  0.0439\, c^2_{t_R}\, m^2_0 \,\nonumber\\
  &-&  0.000119\, c_{M_1}\, c_{M_2}\, M^2\, \cos \Phi_{12} -   0.000501\, c_{M_1}\, c_{M_3}\, M^2\, \cos \Phi_{13} \,\nonumber\\
  &+&  0.000361\, A_0\, c_{A_b}\, c_{M_1}\, M\, \cos \Phi_{1b} -   2.71\, \times\, {10}^{-6}\, A_0\, c_{A_\tau}\, c_{M_1}\,M\, \cos \Phi_{1\tau} \,\nonumber\\
  &-&  0.0000533\, A_0\, c_{A_t}\, c_{M_1}\, M\, \cos \Phi_{1t} - 0.00647\, c_{M_2}\, c_{M_3}\, M^2\, \cos \Phi_{23} \,\nonumber\\
  &+&  0.00427\, A_0\, c_{A_b}\, c_{M_2}\, M\, \cos\Phi_{2b} -  0.0000136\, A_0\, c_{A_\tau}\, c_{M_2}\, M\, \cos \Phi_{2\tau} \,\nonumber\\
  &-&  0.000509\, A_0\, c_{A_t}\, c_{M_2}\, M\, \cos\Phi_{2t} +  0.0172\, A_0\, c_{A_b}\, c_{M_3}\, M\, \cos\Phi_{3b} \,\nonumber\\
  &-&  0.0000363\, A_0\, c_{A_\tau}\, c_{M_3}\, M\, \cos (\Phi_{3\tau} -  0.00205\, A_0\, c_{A_t}\, c_{M_3}\, M\, \cos\Phi_{3t} \,\nonumber\\
  &+&  0.0000222\, A^2_0\, c_{A_b}\, c_{A_\tau}\, \cos \Phi_{b\tau} +  0.00106\, A^2_0\, c_{A_b}\, c_{A_t}\, \cos\Phi_{tb} \,\nonumber\\
  &-&  2.11\, \times\, {10}^{-6}\, A^2_0\, c_{A_\tau}\,c_{A_t}\, \cos \Phi_{t\tau}~,\,\end{eqnarray}
\begin{eqnarray}
m^2_{\tilde \tau_L}&=&0.000011\, A^2_0\, c^2_{A_b} -
  0.00274\, A^2_0\, c^2_{A_\tau} -
  3.11\, \times\, {10}^{-7}\, A^2_0\, c^2_{A_t} +
  0.0365\, c^2_{M_1}\, M^2 \,\nonumber\\
  &+& 0.457\, c^2_{M_2}\,
M^2 + 0.000035\, c^2_{M_3}\, M^2 - 0.0353\, c^2_{H_d}\, m^2_0 +
  0.0325\, c^2_{H_u}\, m^2_0 +
  0.0325\, c^2_{b_R}\, m^2_0 \,\nonumber\\
  &+&
  0.965\, c^2_{\tau_L}\, m^2_0 +
  0.0297\, c^2_{\tau_R}\, m^2_0 +
  0.0325\, c^2_{t_L}\, m^2_0 -
  0.065\, c^2_{t_R}\, m^2_0 \,\nonumber\\
  &-&   0.000204\, c_{M_1}\, c_{M_2}\,M^2\, \cos \Phi_{12} +  3.\, \times\, {10}^{-6}\, c_{M_1}\, c_{M_3}\,M^2\, \cos \Phi_{13} \,\nonumber\\
  &-&   3.46\,    \times\, {10}^{-6}\, A_0\, c_{A_b}\, c_{M_1}\,M\, \cos \Phi_{1b} +  0.00059\, A_0\, c_{A_\tau}\, c_{M_1}\, M\, \cos \Phi_{1\tau} \,\nonumber\\
  &+&   2.42\, \times\, {10}^{-7}\, A_0\, c_{A_t}\, c_{M_1}\,M\, \cos \Phi_{1t} +  0.0000132\, c_{M_2}\, c_{M_3}\,M^2\, \cos \Phi_{23} \,\nonumber\\
  &-&   0.000014\, A_0\, c_{A_b}\, c_{M_2}\, M\, \cos\Phi_{2b} +  0.00162\, A_0\, c_{A_\tau}\, c_{M_2}\, M\, \cos\Phi_{2\tau} \,\nonumber\\
  &+&  1.07\, \times\, {10}^{-6}\, A_0\, c_{A_t}\, c_{M_2}\,M\, \cos \Phi_{2t} -  0.0000176\, A_0\, c_{A_b}\, c_{M_3}\, M\, \cos \Phi_{3b} \,\nonumber\\
  &-&  0.0000178\, A_0\, c_{A_\tau}\, c_{M_3}\, M\, \cos \Phi_{3\tau} +  1.78\, \times\, {10}^{-6}\, A_0\, c_{A_t}\, c_{M_3}\,M\, \cos \Phi_{3t} \,\nonumber\\
  &+&  0.0000222\, A^2_0\, c_{A_b}\, c_{A_\tau}\, \cos \Phi_{b\tau} -  1.28\, \times\, {10}^{-6}\, A^2_0\, c_{A_b}\, c_{A_t}\, \cos \Phi_{tb} \,\nonumber\\
  &-&  1.29\, \times\, {10}^{-6}\, A^2_0\, c_{A_\tau}\,c_{A_t}\, \cos \Phi_{t\tau}~,\,\end{eqnarray}
\begin{eqnarray}
m^2_{\tilde \tau_R}&=&0.0000221\, A^2_0\, c^2_{A_b} -
  0.00548\, A^2_0\, c^2_{A_\tau} -
  6.21\, \times\, {10}^{-7}\, A^2_0\, c^2_{A_t} +
  0.147\, c^2_{M_1}\, M^2 \,\nonumber\\
  &-& 0.00366\, c^2_{M_2}\,
M^2 + 0.0000699\, c^2_{M_3}\, M^2 + 0.0595\, c^2_{H_d}\, m^2_0 -
  0.065\, c^2_{H_u}\, m^2_0 -
  0.065\, c^2_{b_R}\, m^2_0 \,\nonumber\\
  &+&
  0.0595\, c^2_{\tau_L}\, m^2_0 +
  0.929\, c^2_{\tau_R}\, m^2_0 -
  0.065\, c^2_{t_L}\, m^2_0 +
  0.13\, c^2_{t_R}\, m^2_0 \,\nonumber\\
  &-&   0.000409\, c_{M_1}\, c_{M_2}\,M^2\, \cos \Phi_{12} +  5.99\, \times\, {10}^{-6}\, c_{M_1}\, c_{M_3}\,M^2\, \cos \Phi_{13} \,\nonumber\\
  &-&  6.93\, \times\, {10}^{-6}\, A_0\, c_{A_b}\, c_{M_1}\,M\, \cos \Phi_{1b} +  0.00118\, A_0\, c_{A_\tau}\, c_{M_1}\, M\, \cos\Phi_{1\tau} \,\nonumber\\
  &+&  4.83\, \times\, {10}^{-7}\, A_0\, c_{A_t}\, c_{M_1}\,M\, \cos \Phi_{1t} +  0.0000264\, c_{M_2}\, c_{M_3}\,M^2\, \cos \Phi_{23} \,\nonumber\\
  &-&  0.000028\, A_0\, c_{A_b}\, c_{M_2}\, M\, \cos\Phi_{2b} +  0.00324\, A_0\, c_{A_\tau}\, c_{M_2}\, M\, \cos\Phi_{2\tau} \,\nonumber\\
  &+&  2.13\, \times\, {10}^{-6}\, A_0\, c_{A_t}\, c_{M_2}\,M\, \cos \Phi_{2t} -  0.0000352\, A_0\, c_{A_b}\, c_{M_3}\, M\, \cos \Phi_{3b} \,\nonumber\\
  &-&  0.0000355\, A_0\, c_{A_\tau}\, c_{M_3}\, M\, \cos \Phi_{3\tau} +  3.57\, \times\, {10}^{-6}\, A_0\, c_{A_t}\, c_{M_3}\,M\, \cos \Phi_{3t} \,\nonumber\\
  &+&  0.0000444\, A^2_0\, c_{A_b}\, c_{A_\tau}\, \cos \Phi_{b\tau} -  2.55\, \times\, {10}^{-6}\, A^2_0\, c_{A_b}\,c_{A_t}\, \cos \Phi_{tb} \,\nonumber\\
  &-&  2.58\, \times\, {10}^{-6}\, A^2_0\, c_{A_\tau}\,c_{A_t}\, \cos \Phi_{t\tau}~.\,\,\end{eqnarray}
  For Gauginos we found the followings
\begin{equation}
M_1=0.432\, c_{M_1}\, M,\,\,M_2=0.833\, c_{M_2}\, M,\,\,M_3=2.51\,
c_{M_3}\, M,\,.
\end{equation}
Similarly, for trilinear terms we have
\begin{eqnarray}
A_t&=&-0.00198\, A_0\, c_{A_b} +
  3.81\, \times\, {10}^{-6}\, A_0\, c_{A_\tau} +
  0.27\, A_0\, c_{A_t} - 0.0303\, c_{M_1}\, M -
  0.231\, c_{M_2}\, M \,\nonumber\\
  &-& 1.55\, c_{M_3}\, M
\end{eqnarray}
\begin{eqnarray}
A_b&=&0.147\, A_0\, c_{A_b} -
  0.00041\, A_0\, c_{A_\tau} -
  0.0175\, A_0\, c_{A_t} - 0.00484\, c_{M_1}\, M -
  0.0675\, c_{M_2}\, M \,\nonumber\\
  &-& 0.372\, c_{M_3}\,
\end{eqnarray}
\begin{eqnarray}
A_\tau&=&-0.00101\, A_0\, c_{A_b} +
  0.0989\, A_0\, c_{A_\tau} +
  0.0000811\, A_0\, c_{A_t} -
  0.0153\, c_{M_1}\, M - 0.0493\, c_{M_2}\, M \,\nonumber\\
  &+&
  0.00131\, c_{M_3}\, M
\end{eqnarray}
Our expression for $B$ is
\begin{eqnarray}
B&=&B_0 - 0.0095\, A_0\, c_{A_b} -
  0.00276\, A_0\, c_{A_\tau} -
  0.354\, A_0\, c_{A_t} - 0.0301\, c_{M_1}\, M -
  0.371\, c_{M_2}\, M \,\nonumber\\
  &+& 0.518\, c_{M_3}\, M
\end{eqnarray}
and, lastly, for the $\mu$ parameter our result reads
\begin{equation}
\mu=0.995\, \mu_0\,.
\end{equation}

\end{document}